\documentclass[prd,nofootinbib,12pt]{revtex4}
\usepackage[utf8]{inputenc}
\usepackage{lineno}
\usepackage{amsfonts,amssymb,amsmath}
\usepackage{epsfig}
\usepackage{mathrsfs}
\usepackage{amssymb}
\usepackage{graphicx}
\usepackage{amsmath}
\usepackage{xcolor}
\usepackage{color}
\usepackage{appendix}
\usepackage{subcaption}
\usepackage{caption}
\usepackage{ragged2e}
\begin{document}

\title{Krylov Complexity in Non-Inertial Quantum Systems}
\author{Ming-Qi Ma}
 \affiliation{Department of Physics, College of Physics, Mechanical and Electrical Engineering, Jishou University, Jishou 416000, China}
 \author{Shi-Cheng Liu}
 \affiliation{Department of Physics, College of Physics, Mechanical and Electrical Engineering, Jishou University, Jishou 416000, China}
\author{Lei-Hua Liu}%
 \email{liuleihua8899@hotmail.com (corresponding author)}
\affiliation{Department of Physics, College of Physics, Mechanical and Electrical Engineering, Jishou University, Jishou 416000, China}%
\author{Hai-Qing Zhang}
\email{hqzhang@buaa.edu.cn}
\affiliation{%
 Center for Gravitational Physics, Department of Space Science, Beihang University, Beijing,
100191, China}
\affiliation{Peng Huanwu Collaborative Center for Research and Education, Beihang University, Beijing
100191, China
}

\begin{abstract}
{This study formulates observer-dependent Krylov spreading for non-inertial quantum systems driven by linear Bogoliubov transformations. Starting with the closed single Rindler-pair $SU(1,1)$ sector, we show that its Lanczos basis is identical to the Rindler pair-number basis. As a result, the Krylov spread complexity reduces exactly to the mean number of correlated Rindler pairs, $C_k=\vert\beta_k\vert^2$. Within this framework, we demonstrate that Krylov spreading dynamics are governed by the competition between the detuning parameter and the coupling constant, naturally dividing the dynamics into three distinct regimes. Notably, Krylov complexity becomes localized in the detuning-dominated regime. By extending this to a multimode, strictly quadratic Bogoliubov Hamiltonian, we find that inequivalent Rindler wave-packet pairs violate the $C_k=\vert\beta_k\vert^2$ correspondence, thereby highlighting the single-pair $SU(1,1)$ model as an exactly solvable, observer-adapted benchmark. In such multimode scenarios, the mean pair-number eigenstates no longer dictate Krylov complexity. Overall, our work provides a new perspective for analyzing Krylov complexity in non-inertial quantum systems.
 }

\end{abstract}

\maketitle

\section{introduction}
\label{introduction}
Elucidating the spread of quantum information across the causal structure of spacetime remains a fundamental problem at the intersection of quantum gravity and many-body physics \cite{Shenker:2013pqa,Roberts:2014isa,Maldacena:2015waa,Roberts:2016hpo,Swingle:2018ekw}. A robust metric for diagnosing this phenomenon is Krylov complexity, which models the evolution of a simple Heisenberg operator as a wave packet expanding along a one-dimensional Krylov chain \cite{viswanath1994recursion,Parker:2018yvk,Rabinovici:2020ryf,Rabinovici:2022beu,Caputa:2021sib}. While this formalism has been extensively applied to conformal field theories, the SYK model, open quantum systems \cite{Dymarsky:2021bjq,Jian:2020qpp,Bhattacharya:2022gbz,Liu:2022god,Camargo:2022rnt,Adhikari:2022whf,Balasubramanian:2022tpr,Caputa:2022yju}, and cosmological environments \cite{Liu:2025caj,Zhai:2024tkz,Zhai:2024odw,Li:2024iji,Li:2024kfm,Zhai:2025abc,Li:2024ljz}, these investigations predominantly rely on a static observer paradigm. In such setups, the vacuum state, the operator algebra, and the time-evolution generator are rigidly tied to an inertial frame, or at least constrained to a single, predetermined temporal reference.

This assumption, however, becomes highly nontrivial in relativistic quantum field theory. A uniformly accelerated observer naturally expands the field in terms of Rindler modes, attributing a thermal particle content to the Minkowski vacuum via the Unruh effect \cite{Fulling:1972md,Davies:1974th,Hawking:1975vcx,Unruh:1976db,Hollands:2014eia,Takagi:1986kn,Wald:1995hf,Crispino:2007eb,Ford:1997hb}. The field of relativistic quantum information has made extensive use of this phenomenon, demonstrating that acceleration can significantly modify entanglement and quantum communication protocols \cite{Peres:2002wx,Alsing:2003es,Fuentes-Schuller:2004iaz,Alsing:2006cj,Adesso:2007wi,Bruschi:2010mc,Martin-Martinez:2010yva,Friis:2012tb,Ahmadi:2013lxa}. While these analyses are typically formulated using states and reduced density matrices, the Heisenberg perspective invites a slightly different question: If the particle concept and the natural mode decomposition are observer-dependent, should the Krylov description of operator growth inherently depend on the observer as well?

Importantly, we find that the apparent equivalence between Krylov complexity and particle production is a consequence of the closed SU(1,1) structure rather than a fundamental identity of non-inertial quantum dynamics. For an accelerated observer, the Bogoliubov transformation between Minkowski and Rindler modes reorganizes the field into correlated Rindler pair-number sectors. These sectors provide a natural Krylov chain tailored to the non-inertial perspective. We illustrate this construction using a solvable two-mode $SU(1,1)$ model, which is closely related to two-mode squeezing, parametric amplification, and analogue-gravity setups \cite{Caves:1985zz,Schumaker:1985zz,Gerry:1985zz,Yurke:1986ipz,Zhang:1990fy,Hornberger:2012xli,Hu:2018psq,liu2026bogoliubov}.

 In the closed  $SU(1,1)$ sector, we prove that the Lanczos basis coincides exactly with the Rindler pair-number basis. Consequently, the Krylov spread complexity reduces to the mean Rindler pair number,
\begin{equation}
C_K(t)
=
\left\langle \hat N_{\rm pair}\right\rangle_t
=
|\beta_k(t)|^2.
\label{CK-pair-identity-intro}
\end{equation}
We emphasize that this identity is a structural property of the  reduction rather than a generic consequence of linear Bogoliubov dynamics. In this restricted sector, the Krylov entropy and the variance do not constitute independent diagnostics, since the full Krylov probability distribution is geometric and is completely determined by the single parameter $|\beta_k|^2$. Within this exactly solvable benchmark, the competition between the pair-production amplitude $g$ and the detuning parameter $u_2$
separates the dynamics into three regimes. For $|u_2|<|g|$, the Krylov wave packet spreads hyperbolically along the Rindler pair-number chain. At $|u_2|=|g|$, the dynamics exhibits critical growth, whereas for $|u_2|>|g|$, the evolution becomes bounded and oscillatory. In the latter regime, the probability distribution remains concentrated near the low Krylov levels, realizing detuning-induced dynamical confinement in the observer-adapted Krylov chain.
We then go beyond the  approximation while retaining a strictly quadratic Hamiltonian and a linear Bogoliubov transformation. For two inequivalent Rindler wave-packet pairs, each fixed total-pair-number sector contains several orthogonal mode-distribution states. The Lanczos vectors therefore generically cease to be eigenstates of the total pair-number operator. As a
result, the Krylov complexity is no longer determined solely by the mean pair-number eigenstates. This identifies multimode Bogoliubov mixing as the minimal Gaussian mechanism through which Krylov spreading contains dynamical information beyond average Rindler-pair production.

This paper is organized as follows. In Sec.~\ref{Non-Inertial Krylov Framework}, we will establish the observer-dependent Krylov complexity framework for non-inertial quantum systems, deriving the modified wave function and Bogoliubov coefficients. In Sec.~\ref{Acceleration-Controlled Krylov Dynamics}, we will investigate the acceleration-controlled Krylov dynamics and classify them into three distinct regimes. In Sec.~\ref{sec:multimode}, we extend the construction to multimode linear Bogoliubov dynamics and demonstrate the breakdown of the exact relation between Krylov complexity and the mean pair
number. Finally, a brief conclusion and outlook will be presented in Sec.~\ref{conclusion and outlook}.

\section{Non-Inertial Krylov Framework}
\label{Non-Inertial Krylov Framework}

In this section, we formulate the non-inertial framework of Krylov complexity. We begin by introducing the wavefunction relating the Minkowski vacuum to the Rindler vacuum.

\subsection{The wave function}

We first consider a two-mode massless scalar field that appears maximally entangled from an inertial perspective in Minkowski spacetime. Expressed in terms of non-inertial observers (Rindler spacetime) \cite{Fuentes-Schuller:2004iaz}, this state can be described as
\begin{equation}
   \frac{1}{\sqrt{2}} \left( |0_s\rangle^{\mathcal{M}} |0_k\rangle^{\mathcal{M}} + |1_s\rangle^{\mathcal{M}} |1_k\rangle^{\mathcal{M}} \right).
\label{general two mode}
\end{equation}
Here, we consider two observers, Alice and Bob, where Alice only detects mode $s$ while Bob is sensitive exclusively to mode $k$. In this framework, Bob undergoes uniform acceleration, meaning his observations are framed within Rindler spacetime. Consequently, the mode $k$ detected by Bob must be mapped onto the Rindler modes, establishing the transformation between his inertial and non-inertial descriptions. Crucially, the wavefunction connects the Minkowski vacuum to the Rindler vacuum for Bob. For simplicity, we investigate a single wave-packet sector of a scalar field from the perspective of this uniformly accelerated observer. Upon applying the standard wave-packet projection and the  approximation, the Minkowski vacuum naturally expands into a two-mode squeezed state across the left and right Rindler wedges—a well-established consequence of the Fulling-Davies-Unruh effect~ \cite{Fulling:1972md,Davies:1974th,Unruh:1976db}, 
\begin{equation}
    |0_R\rangle^{\mathcal{M}}=\frac{1}{\cosh r_k}\sum_{n=0}^{\infty}\tanh ^n r_k|n_k\rangle_{I}|n_k\rangle_{II},
    \label{two mode squeezd state}
\end{equation}
where $|n_k\rangle_{I}$ and $|n_k\rangle_{II}$ denote the Rindler Fock states in regions $I$ and $II$, respectively, providing the basis over which the Minkowski vacuum $|0_k\rangle^{\mathcal{M}}$ is decomposed. This expansion is intrinsically linked to the KMS condition of the Minkowski vacuum when restricted to a single Rindler wedge \cite{Bisognano:1975ih,Bisognano:1976za}. As demonstrated in Refs.~\cite{Alsing:2003es,Fuentes-Schuller:2004iaz}, this vacuum state is precisely a two-mode squeezed state. The corresponding Bogoliubov transformation relating the Minkowski and Rindler operators is characterized by 
\begin{equation}
    \hat{a}=\alpha_k \hat c_k-\beta_k\hat c_{-k},
    \label{bogoliubov transformation for two mode squeezed state}
\end{equation}
where $\alpha_k=\cosh r_k$ and $\beta_k=\sinh r_k$. Here, $\hat{a}_k$ denotes the annihilation operator for the Minkowski vacuum, while $\hat{b}_k$ represents the annihilation operator for the Rindler vacuum. Note that we have omitted the phase factor conventionally included in standard two-mode squeezed states \cite{Liu:2021nzx,Li:2021kfq,Li:2023ekd}. By inspection, the two-mode squeezed state can be expressed in terms of the Bogoliubov coefficients as follows:
\begin{equation}
  |0_R\rangle^{\mathcal{M}}=\sqrt{1-\left(\frac{\beta_k}{\alpha_k}\right)^2}\sum_{n=0}^{\infty}\left(\frac{\beta_k}{\alpha_k}\right)^n |n_k\rangle_{I}|n_k \rangle_{II}.
  \label{genera wave function}
\end{equation}
When expressed in terms of the Bogoliubov coefficients, this formulation naturally reduces to the two-mode squeezed state \eqref{two mode squeezd state}. Notably, the wavefunction parameterized by these Bogoliubov coefficients \cite{Caves:1985zz,Alsing:2003es,Grishchuk:1990bj} is highly generic and can be identified as a generalized Perelomov coherent state \cite{Craps:2024suj,Perelomov:1986uhd}. Consequently, varying the choice of Bogoliubov coefficients yields distinct wavefunctions. In what follows, we present the detailed derivation of this structure.

We expand the non-inertial vacuum in the Rindler pair-number basis $|n\rangle_K \equiv |n_k, n_{-k}\rangle$:
\begin{equation}
    |0\rangle_R = \sum_{n=0}^{\infty} c_n |n\rangle_K.
    \label{expansion}
\end{equation}
where $c_n$ denotes the coefficient representing the probability amplitude for the corresponding $n$-th wavefunction. By definition, acting with the annihilation operator on the Rindler vacuum yields $\hat{a}|0\rangle_R=0$. Substituting the Bogoliubov transformation \eqref{bogoliubov transformation for two mode squeezed state} and incorporating the expansion \eqref{expansion}, we obtain 
\begin{equation}
    \sum_{n=1}^{\infty} \alpha_k c_n \sqrt{n} |n-1\rangle_K - \sum_{n=0}^{\infty} \beta_k c_n \sqrt{n+1} |n+1\rangle_K = 0.
\end{equation}
By shifting the summation index in the first term (setting $m = n - 1$) and matching the coefficients of each linearly independent basis state $|m\rangle_K$, we obtain the following recurrence relation:
\begin{equation}
    c_{m+1} = \left(\frac{\beta_k(t)}{\alpha_k(t)}\right) c_m.
\end{equation}
This recurrence implies that all amplitudes can be expressed in terms of the vacuum coefficient $c_0$:
\begin{equation}
    c_n = \chi^n c_0, \quad \text{with} \quad \chi = \frac{\beta_k(t)}{\alpha_k(t)}.
\end{equation}
Finally, the coefficient $c_0$ is determined by the normalization condition $\langle 0|0\rangle_R = 1$. The preservation of the commutation relations guarantees that $|\alpha_k(t)|^2 - |\beta_k(t)|^2 = 1$, which ensures that the squared norm satisfies $|\chi|^2 = |\beta_k(t)/\alpha_k(t)|^2 < 1$. Consequently, the convergence of the geometric series is guaranteed: 
\begin{equation}
    1 = |c_0|^2 \sum_{n=0}^{\infty} |\chi|^{2n} = \frac{|c_0|^2}{1 - |\chi|^2}.
\end{equation}
Equipped with these parameters, we can rewrite the generalized Perelomov coherent state in terms of
\begin{equation}
  |0_R\rangle^{\mathcal{M}}=\sqrt{1-|\chi|^2}\sum_{n=0}^{\infty}|\chi|^n |n\rangle_K. 
  \label{genera wave function}
\end{equation}

\subsection{Generalized Bogoliubov transformation}
\label{Generalized Bogoliubov transformation}
As noted in the preceding section, different Bogoliubov transformations yield distinct wavefunctions. In this section, we employ the Hamiltonian formalism to derive the generalized Bogoliubov coefficients, focusing on a Hamiltonian that possesses an $SU(1,1)$ structure. Prior to this derivation, we motivate our choice and explain why this specific structure is adopted.

For the generalized Perelomov coherent state, we define $|n\rangle_K \equiv |n_k, n_{-k}\rangle$, which constitutes a core conceptual innovation of this work. The primary task here is to justify this identification. At first glance, a careful reader will notice that the expansion \eqref{genera wave function} embodies the superposition principle of quantum mechanics, implying that the basis $|n\rangle_K$ is inherently dynamical. Consequently, it is highly natural to let the Hamiltonian act on this Rindler pair-number basis, which we dub the Krylov basis—a choice that warrants a detailed explanation. To determine the generalized Bogoliubov coefficients, one must establish the transformation between these two distinct bases, corresponding to Minkowski and Rindler spacetimes, respectively. The most essential step in this procedure is to calculate the Klein-Gordon inner product, which fundamentally defines the space of classical solutions, 
\begin{equation}
\Omega(\phi_1,\phi_2) = -i(\phi_1,\phi_2)_{KG},
\label{KG production}
\end{equation}
For the mode decomposition in Minkowski spacetime, we write $\phi = \int d\omega \left(f_\omega \hat a_\omega + f_\omega^* \hat a_\omega^\dagger \right)$, whereby the phase space becomes a symplectic vector space spanned by the Darboux coordinates $(\hat a_\omega,\hat a_\omega^\dagger)$. A Bogoliubov transformation connecting these two vacua is a linear canonical transformation that preserves the symplectic form \cite{Caves:1981hw}, 
\begin{equation}
S \in Sp(2\infty,\mathbb{R}),
\label{Sp infinity}
\end{equation}
where $\infty$ denotes the infinite degrees of freedom in quantum field theory (QFT), which simultaneously guarantees the preservation of the canonical commutation relations. Upon applying the wave-packet projection and  truncation, this transformation reduces to an effective two-dimensional symplectic sector,
\begin{equation}
S_k \in Sp(2,\mathbb{R}) \simeq SU(1,1),
\end{equation}
which generates the effective two-mode squeezing dynamics foundational to the Krylov construction. Ultimately, the $SU(1,1)$ structure emerges directly from the Klein-Gordon (KG) symplectic form, which physically motivates our adoption of a Hamiltonian governed by the $SU(1,1)$ group structure. Furthermore, to accommodate this generalized Bogoliubov transformation, it is necessary to introduce a generalized Rindler retarded time $u$:
\begin{equation}
U = p(u),
\end{equation}
where the standard uniform acceleration is recovered by setting $p(u) = -\frac{1}{a} e^{-a u}$. Here, we have focused on the null sector of a massless scalar field defined by the Minkowski null coordinate $U = t - z$. The corresponding Minkowski and generalized Rindler positive-frequency modes are given by 
\begin{align}
f_\omega(U) &= \frac{1}{\sqrt{4 \pi \omega}} e^{-i \omega U}, \quad \omega > 0, \\
v_\Omega(u) &= \frac{1}{\sqrt{4 \pi \Omega}} e^{-i \Omega u}, \quad \Omega > 0.
\end{align}
The generalized Rindler modes can be expanded in Minkowski modes via Bogoliubov kernels:
\begin{equation}
v_\Omega(u) = \int_0^\infty d\omega \Big[ A_{\Omega \omega}[p] f_\omega(p(u)) + B_{\Omega \omega}[p] f_\omega^*(p(u)) \Big],
\end{equation}
with
\begin{equation}
A_{\Omega \omega}[p] = (f_\omega, v_\Omega)_{KG}, \quad
B_{\Omega \omega}[p] = -(f_\omega^*, v_\Omega)_{KG}.
\end{equation}
Projecting onto a narrow Rindler wave-packet yields  coefficients:
\begin{align}
\alpha_k[p] &= N_k[p] \int d\Omega\, d\omega\, F_k^*(\Omega) A_{\Omega \omega}[p] G_k(\omega), \\
\beta_k[p] &= N_k[p] \int d\Omega\, d\omega\, F_k^*(\Omega) B_{\Omega \omega}[p] G_k^*(\omega),
\label{general bogoliubov transformation}
\end{align}
where the normalization factor is
\begin{equation}
N_k[p] = \frac{1}{\sqrt{|\alpha_k[p]|^2 - |\beta_k[p]|^2}}, \quad |\alpha_k[p]|^2 - |\beta_k[p]|^2 = 1.
\end{equation}
These coefficients reduce to the standard Unruh squeezing parameters in the stationary limit, forming the basis for constructing the  Krylov description used in the main text. Finally, upon applying the wave-packet projection and  truncation, the generalized Bogoliubov transformation assumes the same structural form as Eq.~\eqref{bogoliubov transformation for two mode squeezed state}. 

Next, we utilize the generalized Lanczos algorithm \cite{Bhattacharjee:2022lzy,Bhattacharya:2023zqt} to demonstrate that the Rindler pair-number sector naturally forms the Krylov basis. As previously noted, the generalized Bogoliubov transformation possesses an $SU(1,1)$ algebraic framework. Alternatively, this Bogoliubov transformation can be derived from a Hamiltonian that shares the same $SU(1,1)$ group structure, expressed as follows:
\begin{equation}
    H=2u_2K_0+g(K_+ + K_-),
    \label{total hamiltonian}
\end{equation}
where $K_+=\hat c_k^\dagger \hat c_{-k}^\dagger$, $K_-=\hat c_k \hat c_{-k}$, and $K_0=\frac{1}{2}\left(\hat c_k^\dagger \hat c_k+\hat c_{-k}\hat c_{-k}^\dagger\right)$, with $g$ denoting the pair-production amplitude and $u_2$ representing the detuning parameter. When acting on the Krylov basis $|n\rangle_K$, these $SU(1,1)$ generators satisfy 
\begin{align}
K_+|n\rangle_K&=(n+1)|n+1\rangle_K,\\
K_-|n\rangle_K&=n|n-1\rangle_K,\\
K_0|n\rangle_K&=\left(n+\frac12\right)|n\rangle_K .
\end{align}
Therefore,
\begin{equation}
H|n\rangle_K
=
g(n+1)|n+1\rangle_K
+
2u_2\left(n+\frac12\right)|n\rangle_K
+
gn|n-1\rangle_K .
\label{eq:tridiagonal_chain}
\end{equation}
Eq.~\eqref{eq:tridiagonal_chain} constitutes the central structural result of this section, demonstrating that the Rindler pair-number basis maps onto a semi-infinite tridiagonal Krylov chain. In the standard Lanczos notation, this correspondence is manifested as 
\begin{equation}
H|n\rangle_K=b_{n+1}|n+1\rangle_K+a_n|n\rangle_K+b_n|n-1\rangle_K ,
\label{lanczos algorithm}
\end{equation}
we obtain
\begin{equation}
a_n=2u_2\left(n+\frac12\right),
\qquad
b_n=|g|n ,
\qquad
b_0=0 .
\label{eq:Lanczos_coefficients}
\end{equation}
Thus, $g$ controls the hopping amplitude along the Krylov chain, while $u_2$ produces a level-dependent onsite detuning. These calculations establish an effective Krylov-chain construction tightly confined within the closed two-mode $SU(1,1)$ sector. Having clarified these foundational concepts, we now proceed to formulate the explicit method for evaluating the generalized Bogoliubov transformation.

Recalling that the standard two-mode squeezed state is conventionally generated by acting the squeezing operator $\mathcal{S}=\exp (-\eta\hat{c}^\dagger_{-k}\hat{c}^\dagger_k+\bar\eta\hat c_k\hat{c}_{-k})$ (with $\eta \in \mathbb{C}$) on the vacuum, Ref.~\cite{Zhai:2025abc} establishes that the generator $-\eta\hat{c}^\dagger_{-k}\hat{c}^\dagger_k+\bar\eta\hat c_k\hat{c}_{-k}$ belongs precisely to the closed sector of the Hamiltonian \eqref{total hamiltonian} under the parameterization $\eta=r_k\exp(2i \phi_k)$. In this work, we extend this squeezing operator into a generalized displacement operator $\mathcal{S}=\exp(i Ht)$ \cite{Zhai:2025abc}, where $H$ denotes the full Hamiltonian \eqref{total hamiltonian}. Utilizing the canonical bosonic commutation relations $ [\hat c_i,\hat c_j^\dagger]=\delta_{ij}$, we obtain $[\hat c_k,H]=u_2\hat c_k+g\hat c_{-k}^\dagger$ and $[\hat c_{-k}^\dagger,H]=-u_2\hat c_{-k}^\dagger-g\hat c_k$. For convenience, we introduce the two-component operator: 
\begin{equation}
    \mathbf C
    =
    \begin{pmatrix}
        \hat c_k\\
        \hat c_{-k}^\dagger
    \end{pmatrix},
\end{equation}
these commutators can be written as
\begin{equation}
    [\mathbf C,H]=\mathcal M\mathbf C,
    \qquad
    \mathcal M=
    \begin{pmatrix}
        u_2 & g\\
        -g & -u_2
    \end{pmatrix}.
\end{equation}
In the main text, we use the in-operator convention $ \mathbf A_{\rm in}(t)=e^{-iHt}\mathbf C e^{iHt}$ with $A_{\rm in}(t)=(\hat{a}_k,\hat{a}_{-k} )$. After some algebra, we could derive 
\begin{equation}
    i\frac{d}{dt}\mathbf A_{\rm in}(t)
    =
    -\mathcal M\,\mathbf A_{\rm in}(t),
\end{equation}
and therefore we also obtain $\mathbf A_{\rm in}(t)=e^{i\mathcal M t}\mathbf C $ \cite{Lewis:1968tm,Parker:1969au,Guth:1985ya,Guven:1987bx}. This choice of convention yields the Bogoliubov coefficients utilized throughout this work. If one instead adopts the standard Heisenberg picture convention, $\mathbf A_H(t)=e^{iHt}\mathbf C e^{-iHt}$, the signs of the imaginary parts of $\alpha_k$ and $\beta_k$ are reversed. Crucially, the physical observables, such as $|\beta_k|^2$, remain invariant. The matrix $\mathcal M$ satisfies the relation $\mathcal M^2 =(u_2^2-|g|^2)\mathbb I$. By defining 
\begin{equation}
    \Omega^2=(|g|^2-u_2^2),
\end{equation}
which can be identified as the modified frequency of the Rindler spacetime; simultaneously, the exponential matrix can be explicitly evaluated as 
\begin{equation}
    e^{i\mathcal M t}
    =
    \cosh\Omega t\times\mathbb I
    +
    i\mathcal M \frac{\sinh\Omega t}{\Omega}.
\end{equation}
Substituting the explicit matrix \(\mathcal M\), we explicitly obtain
\begin{equation}
    e^{i\mathcal M t}
    =
    \begin{pmatrix}
        \cosh\Omega t
        +
        iu_2\dfrac{\sinh\Omega t}{\Omega}
        &
        ig\dfrac{\sinh\Omega t}{\Omega}
        \\[2mm]
        -ig\dfrac{\sinh\Omega t}{\Omega}
        &
        \cosh\Omega t
        -
        iu_2 \dfrac{\sinh\Omega t}{\Omega}
    \end{pmatrix}.
\end{equation}
Thus, we could have, 
\begin{equation}
    \alpha_k(t)
    =
    \cosh\Omega t
    +
    iu_2\frac{\sinh\Omega t}{\Omega},
    \qquad
    \beta_k(t)
    =
    ig\frac{\sinh\Omega t}{\Omega},
    \label{generalized bogoliubov coefficient}
\end{equation}
where $t$ denotes the temporal parameter, explicitly verifying the normalization condition $|\alpha_k|^2-|\beta_k|^2=1$. Physically, the system can be classified into three distinct regimes: the supercritical case $|g|>u_2$, the critical case $|g|=u_2$, and the subcritical case $|g|<u_2$. This classification leads to structurally different behaviors for the Krylov complexity, which will be thoroughly investigated in the subsequent sections. In the limit of vanishing detuning ($u_2=0$), this formalism elegantly reduces to the standard two-mode squeezed state. Consequently, the coefficients simplify to $\alpha_k(t)=\cosh(gt)$ and $\beta_k(t)=i\sinh(gt)$, exactly reproducing the conventional Unruh effect \cite{Unruh:1976db}, where $gt$ explicitly assumes the role of the Unruh squeezing parameter $r_k$. This correspondence naturally establishes an acceleration-time dictionary $gt=r_k(a)$, defined by $r_k(a)=\frac12\ln\!\left[\coth\left(\frac{\pi\omega_k}{2a}\right)\right]$, where $a$ is the proper acceleration and $\omega_k$ is the Rindler mode frequency. Ultimately, this demonstrates a clear physical interplay: the proper acceleration defines the external non-inertial background, whereas the detuning parameter $u_2$ rigidly governs the internal Krylov dynamics. 

\subsection{Krylov complexity}
Finally, a pivotal physical quantity of interest is the Krylov complexity, which is formally defined as 
\begin{equation}
 C_K(t)=\sum_{n=0}^{\infty} n |\varphi_n(t)|^2,
 \label{definition of krylov complexity}
\end{equation}
where $\varphi_n(t)$ denotes the probability amplitude at the $n$-th Krylov site. Crucially, by evaluating the Krylov complexity using the generalized Bogoliubov coefficients \eqref{generalized bogoliubov coefficient}, we arrive at the central result of this paper: 
\begin{equation}
    C_K(t)=|\beta_k(t)|^2.
    \label{krylov complexity}
\end{equation}
Equation~\eqref{krylov complexity} is a direct corollary of the exact basis correspondence established above. Since the Lanczos level $n$ coincides with the number of correlated Rindler pairs in the closed  sector, one obtains
\begin{equation}
C_K(t)
=
\sum_{n=0}^{\infty}n|\phi_n(t)|^2
=
\left\langle \hat N_{\rm pair}\right\rangle_t
=
|\beta_k(t)|^2.
\end{equation}
A closely related correspondence between particle production and Krylov complexity has also been discussed in
Ref.~\cite{Li:2026oxy,Adhikari:2022oxr}. The  model therefore serves as an exactly solvable observer-adapted benchmark. Its importance is that it provides a controlled limit in which the relation between observer-dependent particle production and Krylov spreading can be analytically established. The deviation from this limit reveals the genuinely new information encoded by the Krylov construction. Rather, it identifies the precise condition under which Krylov spreading reduces to a conventional particle-production observable: the Hamiltonian must
remain confined to a single tridiagonal $SU(1,1)$ pair ladder. The  model therefore serves as an exactly solvable observer-adapted benchmark. In Sec.~IV, we show that this reduction generically fails in a multimode linear Bogoliubov system, where the Lanczos vectors no longer possess a definite total pair number.

\section{Acceleration-Controlled Krylov Dynamics}
\label{Acceleration-Controlled Krylov Dynamics}
We now investigate the observer-dependent Krylov dynamics. Guided by the standard Unruh effect, we fix the external squeezing parameter $r_k(a)$. By introducing the dimensionless parameter $\eta = u_2 / g$, the dynamical phase can be rewritten as $\Omega t = \sqrt{1 - \eta^2} r_k$, where $r_k = gt$. Consequently, the Krylov dynamics can be classified into three distinct regimes: $\eta < 1$, $\eta = 1$, and $\eta > 1$. Accordingly, the Krylov complexity manifests in three distinct analytical forms corresponding to these regimes. 
\begin{figure}[htbp]
    \centering

    \begin{subfigure}{0.48\textwidth}
        \centering
        \includegraphics[width=\textwidth]{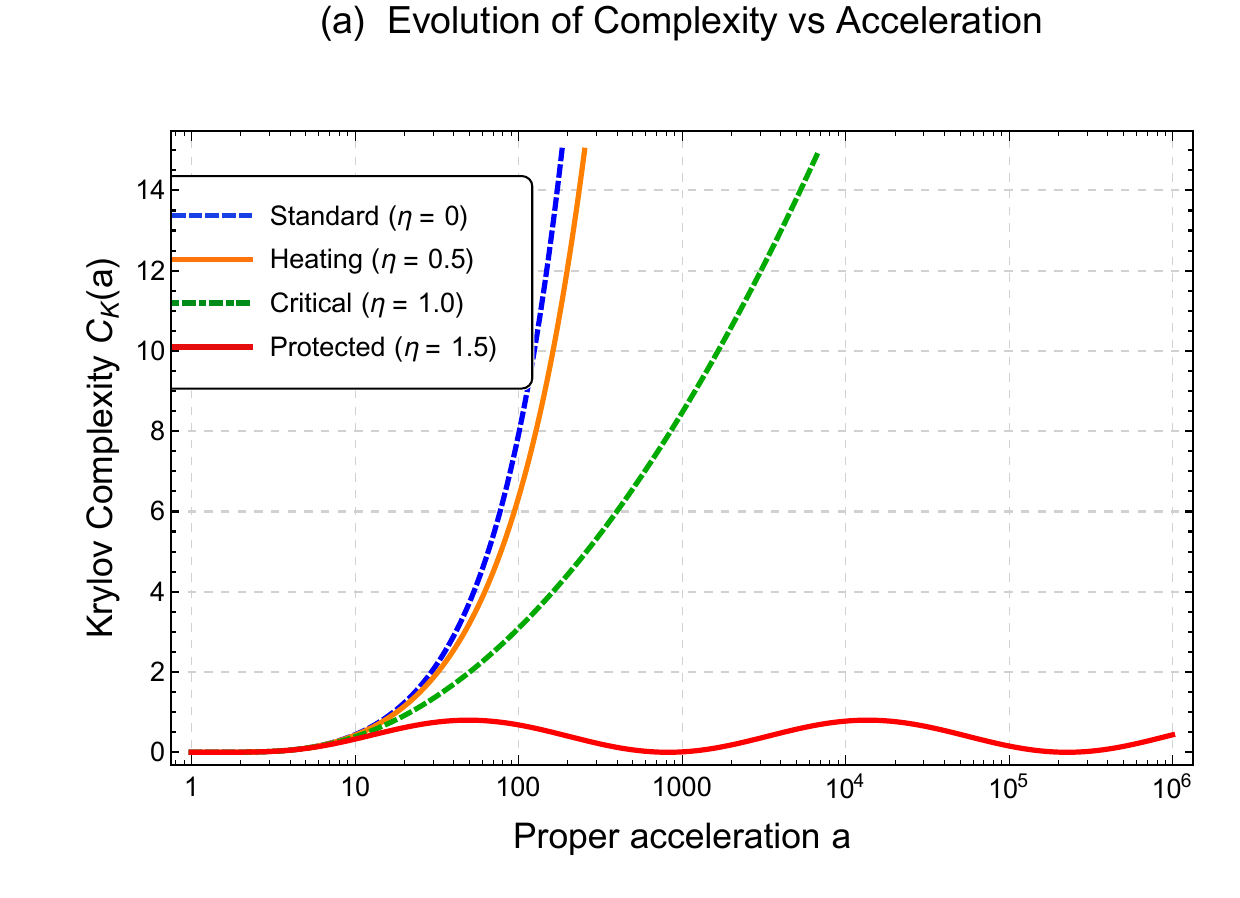}
        \label{fig:complexity}
    \end{subfigure}
    \hfill
    \begin{subfigure}{0.48\textwidth}
        \centering
        \includegraphics[width=\textwidth]{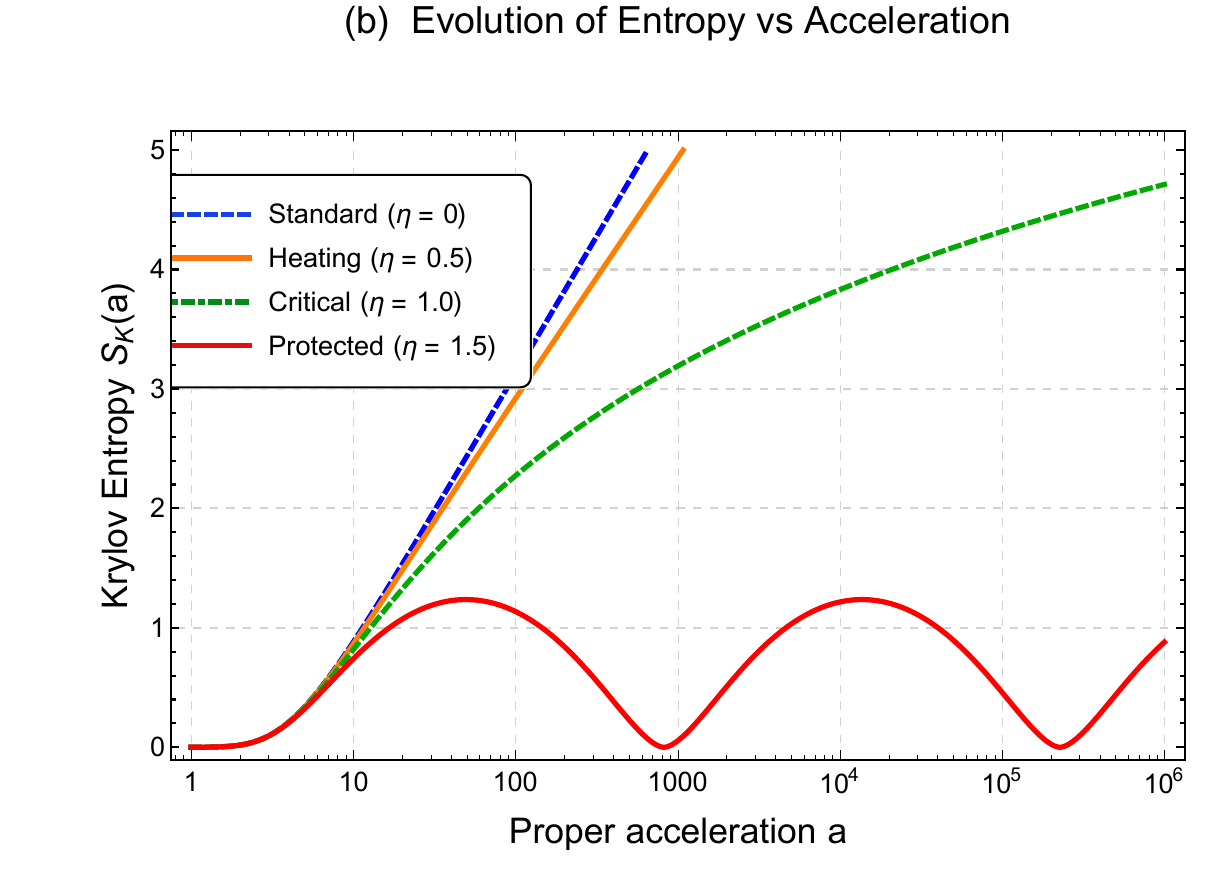}
        \label{fig:entropy}
    \end{subfigure}

    \caption{\justifying
    Acceleration-controlled Krylov dynamics as a function of the dimensionless detuning $\eta = u_2/g$. Left panel: Krylov complexity $C_K(a)$ versus the proper acceleration $a$. Right panel: Krylov entropy $S_K(a)$ versus the proper acceleration $a$. Both plots are generated using $\eta \in (0,0.5,1,1.5)$ and $g = 1$, where the discriminant $\Omega^2 = g^2(1 - \eta^2)$ separates hyperbolic spreading ($|\eta| < 1$), critical growth ($|\eta| = 1$), and bounded motion ($|\eta| > 1$) in the Krylov space. The subcritical detuning regime ($|\eta| > 1$) clearly manifests the dynamical confinement of both Krylov complexity and entropy.}
    \label{fig:comparison}
\end{figure}

Without loss of generality, we set $g=1$. The first case corresponds to the weak-detuning (supercritical) regime where $\eta < 1$. In this regime, the Krylov complexity can be explicitly evaluated as 
\begin{equation}
    C_{K}(a)=\frac{1}{1-\eta^2}\sinh^2\bigg[\sqrt{1-\eta^2}r(a)\bigg], 
    \label{krylov complexity as eta smaller than one}
\end{equation}
This regime encapsulates acceleration-enhanced operator spreading. As the proper acceleration increases, the operator wave packet penetrates deeper into the Rindler-Bogoliubov pair-number Krylov chain. This underlying mechanism perfectly accounts for the rapid growth exhibited by the standard ($\eta=0$) and weak-detuning curves in Fig.~\ref{fig:comparison}. While detuning suppresses the growth rate relative to the zero-detuning baseline, it remains insufficient to halt the hyperbolic expansion. Upon reaching the critical threshold $|\eta|=1$ (corresponding to $u_2=1$ in our setup), the Krylov complexity simplifies to: 
\begin{equation}
    C_K=r^2(a),
    \label{critical krylov complexity}
\end{equation}
This marks the boundary between unbounded operator growth and bounded Krylov motion. The critical curve in Fig.~\ref{fig:comparison} grows at a significantly lower rate than the un-detuned trajectories, indicating that the detuning has begun to suppress resonant propagation along the Krylov chain, although complete localization is not yet achieved. Throughout the regime $|\eta|\leq 1$, the Krylov complexity diverges indefinitely over time. This behavior stands in stark contrast to the findings in Ref.~\cite{Bhattacharjee:2022lzy}, where the Krylov complexity initially grows exponentially before saturating to a constant value. This underscores our first key result: the Krylov complexity of the Rindler pair chain grows indefinitely in the weak-detuning regime, without ever exhibiting saturation. 

Once $|\eta|>1$, then the Krylov complexity will become 
\begin{equation}
    C_K=\frac{1}{\eta^2-1}\sin^2\bigg[\sqrt{\eta^2-1}r_k(a)\bigg].
    \label{krylov complexity bigger than critical value}
\end{equation}
In this regime, $\Omega$ becomes purely imaginary, driving a fundamental transition in the Bogoliubov evolution from hyperbolic to trigonometric dynamics—a behavior clearly visualized in Fig.~\ref{fig:comparison}. This transition leads to our most critical result: the Krylov complexity of the Rindler pair chain undergoes strict localization driven by the strong detuning. Leveraging the inherently bounded nature of trigonometric functions, we establish a rigorous upper bound, $C_K\leq \frac{1}{\eta^2-1}$. This explicitly bounded complexity provides a definitive mathematical signature of Krylov localization within this domain of the parameter space. 

We next examine the Krylov entropy as a probability-level characterization of the same  Krylov distribution. The Krylov entropy is formally defined as 
\begin{equation}
S_K(t)
=
-\sum_{n=0}^{\infty}P_n(t)\ln P_n(t).
\end{equation}
where $P_n=|\phi_n|^2$ represents the probability distribution at the $n$-th Krylov site, with $\phi_n=\sqrt{1-|\chi|^2}|\chi|^n$ denoting the probability amplitude of the generalized coherent state which serves as a diagnostic tool for Krylov localization \cite{brouder2007exponential}. Consequently, the probability simplifies to the geometric distribution $P_n=(1-|\chi|^2)|\chi|^{2n}$. Utilizing the relation $|\chi|^2=\frac{C_K}{C_K+1}$ derived from the definition of Krylov complexity \eqref{krylov complexity}, the Krylov entropy can be explicitly evaluated as follows: 
\begin{equation}
    S_K=(1+C_K)\ln(1+C_K)-C_K\ln C_K.
    \label{krylov entropy}
\end{equation}
Accordingly, within the closed  sector, $S_K$ does not represent an independent diagnostic beyond $C_K$ or the geometric pair-number distribution. It is retained here as a probability-level characterization of the same exactly solvable Krylov wave packet and as a benchmark for the multimode extension considered in Sec.~IV. Thus, it is completely determined by the Krylov complexity,  it naturally inherits the same three-regime structure. As illustrated in Fig.~\ref{fig:comparison}, the entropy grows rapidly in the heating regime, exhibits slower growth at the critical boundary, and remains strictly bounded in the protected regime. The probability-level origin of this bounded entropy production will be clarified in the next section.  

The central message conveyed by Fig.~\ref{fig:comparison} is thus that acceleration alone does not uniquely determine the fate of information scrambling. Instead, the proper acceleration $a$ and the internal detuning $u_2$ jointly govern the Krylov dynamics. Weak detuning allows the Unruh-induced Bogoliubov mixing to drive unbounded operator growth, whereas strong detuning converts this growth into bounded Krylov oscillations. In this sense, the protected regime realizes a non-inertial form of Krylov saturation. Consequently, the suppression of Krylov-entropy production follows directly from the robust confinement of the operator wave packet within the Krylov space.

We already shown the Krylov complexity is confined in the detuning regime, especially for $\eta\gg 1$. Thus, one may ask wether the Krylov dynamical chain can be localized or not. In this section, we will provide a new mechanism of localization for Krylov spreading. The $P_n=|\phi_n|^2$ denotes the wave packet for specific wave function dubbed as a diagnostic tool for Krylov localization \cite{brouder2007exponential}. Based on investigations in Sec.~\ref{Acceleration-Controlled Krylov Dynamics}, the distribution can be expressed in terms of Krylov complexity as follows, 
\begin{equation}
    P_n(a)
    =
    \frac{1}{1+C_K(a)}
    \left[
    \frac{C_K(a)}{1+C_K(a)}
    \right]^n .
    \label{krylov distribution}
\end{equation}
Here, we introduce another key physical quantity: the variance of $P_n(a)$, which characterizes the width of the wave packet and describes its spreading behavior along the Krylov chain.
\begin{equation}
    \Delta n^2=\langle n^2\rangle-\langle n\rangle^2=C_K(1+C_K),
    \label{variance of distribution}
\end{equation}
meanwhile, the expectation value of this geometric distribution is given by $\langle n\rangle = C_K$. For the same reason, the variance
$\Delta n^2=C_K(1+C_K)$ is not independent of the mean Krylov position. Both quantities are fixed by the single parameter $|\beta_k|^2$ in the closed $SU(1,1)$ sector.

\begin{figure}[htbp]
    \centering

    \begin{subfigure}{0.6\textwidth}
        \centering
        \includegraphics[width=\textwidth]{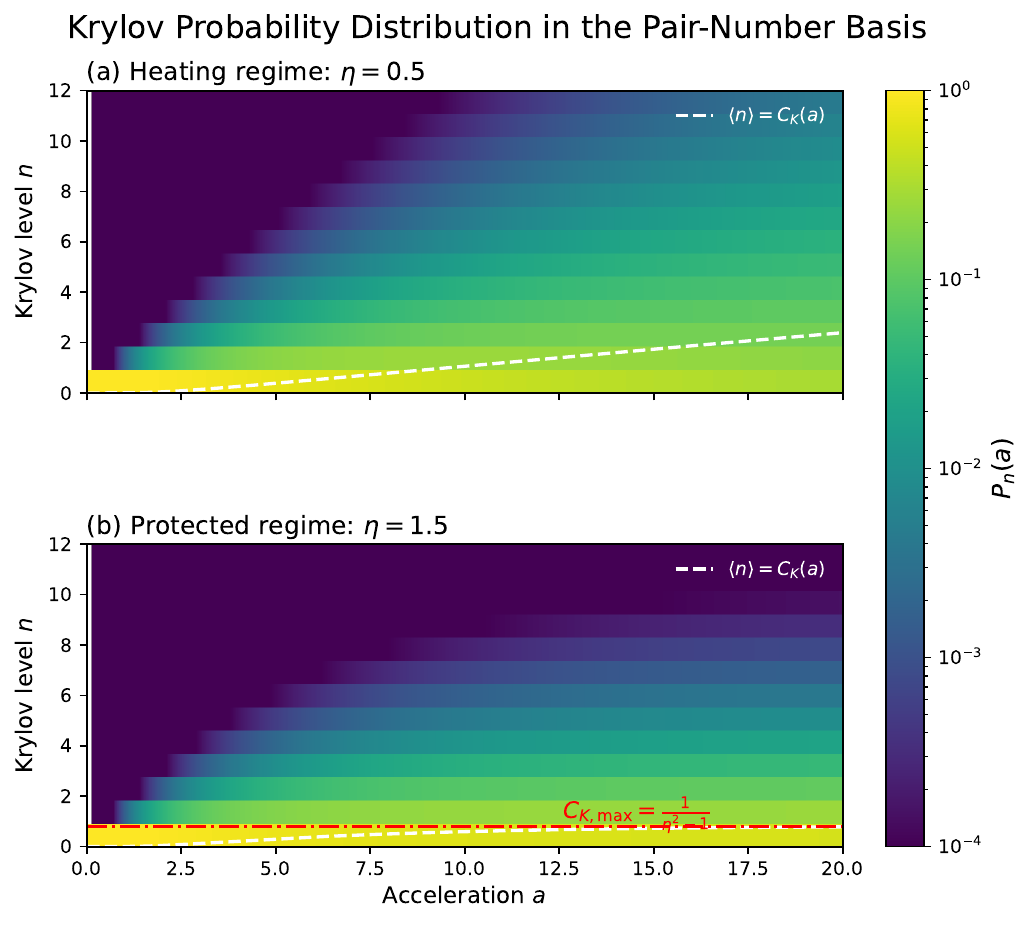}
        \label{fig:distribution}
    \end{subfigure}

    \caption{\justifying
    Probability-level signature of Krylov localization. The color scale illustrates the probability distribution $P_n(a)$ over the pair-number Krylov chain for representative dimensionless detunings $\eta=u_2/g$. The dashed curve tracks the mean Krylov position $\langle n\rangle=C_K(a)$. In the supercritical (weak-detuning) regime $\eta=0.5$, the distribution broadens toward higher Krylov levels as the acceleration increases, serving as a direct hallmark of operator-space delocalization. Conversely, in the subcritical (strong-detuning) regime $\eta=1.5$, the distribution remains tightly confined to lower Krylov levels, with the mean position strictly bounded by $C_{K,\max}=\frac{1}{\eta^2-1}$, thereby demonstrating detuning-induced Krylov localization. }
    \label{fig:distribution}
\end{figure}

Fig.~\ref{fig:distribution} clearly illustrates this underlying localization mechanism. In the heating regime, increasing acceleration inherently broadens the distribution towards higher Krylov levels. Conversely, in the protected regime, the wave packet remains strictly confined to low Krylov levels, with its mean position $\langle n\rangle=C_K(a)$ rigidly constrained below the theoretical upper bound $C_{K,\max}$ denoted in the red solid-dashed line. Ultimately, this probability-level confinement serves as the definitive signature of detuning-induced Krylov localization.

Finally, we observe that in the limit of vanishing acceleration ($a \to 0$), the observer-dependent Krylov complexity identically vanishes, as is manifest from the exact relation $C_K = \sinh^2 r_k$. This behavior does not contradict the conventional Krylov complexity evaluated in standard inertial systems; indeed, within our framework, we have explicitly adopted the Rindler Fock states as the reference vacuum. This apparent discrepancy is naturally resolved by the well-established paradigm that Krylov complexity intrinsically depends on the choice of the initial state \cite{Parker:2018yvk}.

\section{BEYOND THE SINGLE-PARTICLE SU(1,1) SECTOR: MULTIMODE BOGOLIUBOV KRYLOV DYNAMICS}
\label{sec:multimode}

The exact identity established in the preceding sections,
\begin{equation}
C_K(t)=\left\langle \hat N_{\rm pair}\right\rangle_t,
\end{equation}
follows from the fact that, in the closed  $SU(1,1)$
sector, each scalar Lanczos vector is a pair-number eigenstate whose
eigenvalue is equal to its Krylov level. In this section, we show that
this correspondence is not a generic consequence of linear
Bogoliubov dynamics. It is already broken by the minimal extension
to two inequivalent Gaussian pair-production channels. Importantly,
the Hamiltonian remains strictly quadratic throughout this section;
no nonlinear or non-Gaussian interaction is introduced.

\subsection{Multimode quadratic Bogoliubov Hamiltonian}
\label{subsec:multimode-H}

Let
\begin{equation}
\hat{\boldsymbol c}_{I}
=
\begin{pmatrix}
\hat c_{1,I}\\
\hat c_{2,I}
\end{pmatrix},
\qquad
\hat{\boldsymbol c}_{II}
=
\begin{pmatrix}
\hat c_{1,II}\\
\hat c_{2,II}
\end{pmatrix}
\end{equation}
denote two normalized Rindler wave-packet modes in wedges $I$ and
$II$. A general quadratic Hamiltonian containing number-conserving
and pair-production terms can be written as
\begin{align}
H_{\rm G}
={}&
\hat{\boldsymbol c}_{I}^{\dagger}
A_I
\hat{\boldsymbol c}_{I}+\hat{\boldsymbol c}_{II}^{\dagger}
A_{II}
\hat{\boldsymbol c}_{II}
+
\left(
\hat{\boldsymbol c}_{I}^{\dagger}
G
\hat{\boldsymbol c}_{II}^{\dagger\,T}
+
\hat{\boldsymbol c}_{II}^{T}
G^{\dagger}
\hat{\boldsymbol c}_{I}
\right),
\label{general-multimode-H}
\end{align}
where $A_I$ and $A_{II}$ are Hermitian matrices and $G$ is the
multimode pair-production matrix. Introducing the Nambu operator
\begin{equation}
\hat{\boldsymbol{\mathcal C}}
=
\begin{pmatrix}
\hat{\boldsymbol c}_{I}\\
\hat{\boldsymbol c}_{II}^{\dagger}
\end{pmatrix},
\end{equation}
the Heisenberg equations are linear,
\begin{equation}
i\frac{d}{dt}\hat{\boldsymbol{\mathcal C}}
=
\mathcal M_{\rm BdG}
\hat{\boldsymbol{\mathcal C}},
\qquad
\mathcal M_{\rm BdG}
=
\begin{pmatrix}
A_I & G\\
-G^{*} & -A_{II}^{T}
\end{pmatrix}.
\label{multimode-BdG}
\end{equation}
Consequently,
\begin{equation}
\hat{\boldsymbol{\mathcal C}}(t)
=
e^{-i\mathcal M_{\rm BdG}t}
\hat{\boldsymbol{\mathcal C}}(0),
\end{equation}
which is a linear, matrix-valued Bogoliubov transformation.
The pair-production matrix admits a singular-value decomposition. After passive unitary rotations of the wave-packet modes, its minimal two-channel normal form is characterized by two real, non-negative singular values $g_1$ and $g_2$. We therefore consider the quadratic Hamiltonian
\begin{equation}
H_{\rm mm}
=
\sum_{\mu=1}^{2}
\left[
2u_{\mu}K_0^{(\mu)}
+
g_{\mu}
\left(
K_+^{(\mu)}+K_-^{(\mu)}
\right)
\right],
\label{two-channel-H}
\end{equation}
where $K_+^{(\mu)}=\hat c_{\mu,I}^{\dagger}\hat c_{\mu,II}^{\dagger}$, $K_-^{(\mu)}=\hat c_{\mu,I}\hat c_{\mu,II}$ and $K_0^{(\mu)}=\frac{1}{2}\left(
\hat c_{\mu,I}^{\dagger}\hat c_{\mu,I}+\hat c_{\mu,II}\hat c_{\mu,II}^{\dagger}\right)$. The two channel algebras commute with each other and satisfy
\begin{equation}
[K_0^{(\mu)},K_{\pm}^{(\nu)}]
=
\pm\delta_{\mu\nu}K_{\pm}^{(\mu)},
\qquad
[K_+^{(\mu)},K_-^{(\nu)}]
=
-2\delta_{\mu\nu}K_0^{(\mu)}.
\end{equation}
For each channel, we have $\Omega_{\mu}^{2}=g_{\mu}^{2}-u_{\mu}^{2}$ and $\beta_{\mu}(t)=i g_{\mu}\frac{\sinh(\Omega_{\mu}t)}{\Omega_{\mu}}$, with the usual trigonometric continuation when
$\Omega_{\mu}^{2}<0$. The total number of correlated pairs is defined by
\begin{equation}
\hat N_{\rm pair}
=
\sum_{\mu=1}^{2}
\hat c_{\mu,I}^{\dagger}\hat c_{\mu,I},
\label{total-pair-number}
\end{equation}
which is equal to the occupation number in wedge $II$ on the paired subspace. Its expectation value is $\left\langle \hat N_{\rm pair}\right\rangle_t=\sum_{\mu=1}^{2}
|\beta_{\mu}(t)|^{2}$. For a scalar Lanczos construction, we take $|\Psi(t)\rangle=\sum_{n=0}^{\infty} \phi_n(t)|K_n\rangle$, and then
the two quantities of interest take the general forms
\begin{align}
C_K(t)
&=
\sum_{n=0}^{\infty}
n|\phi_n(t)|^{2},
\label{multimode-CK-definition}
\\
\left\langle \hat N_{\rm pair}\right\rangle_t
&=
\sum_{m,n=0}^{\infty}
\phi_m^{*}(t)\phi_n(t)
\left\langle K_m\left|
\hat N_{\rm pair}
\right|K_n\right\rangle.
\label{multimode-N-definition}
\end{align}
Therefore, the exact identity
$C_K=\langle\hat N_{\rm pair}\rangle$ follows only when
\begin{equation}
\left\langle K_m\left|
\hat N_{\rm pair}
\right|K_n\right\rangle
=
n\delta_{mn}
\label{number-Krylov-condition}
\end{equation}
throughout the cyclic Krylov subspace. Equation
\eqref{number-Krylov-condition} is automatically satisfied in the
closed  ladder, but it is not generic in a multimode
system.

\subsection{Breakdown induced by inequivalent detunings}
\label{subsec:unequal-detuning}
In this section, we will give a detailed example to illustrate the breakdown of the $C_K=|\beta_k|^2$ for the multi-mode case. We introduce the normalized two-channel pair-number basis $|n_1,n_2\rangle\equiv|n_1\rangle_{1,I}|n_1\rangle_{1,II}
\otimes|n_2\rangle_{2,I}|n_2\rangle_{2,II}$ with $\hat N_{\rm pair}|n_1,n_2\rangle= (n_1+n_2)|n_1,n_2\rangle$. The additive constant $u_1+u_2$ in Eq.~\eqref{two-channel-H} produces only a global phase. We thus use the shifted Hamiltonian $\widetilde H_{\rm mm}= H_{\rm mm}-(u_1+u_2)\mathbf 1$, where it will not change the Krylov probabilities or the spread complexity. Starting from $|K_0\rangle=|0,0\rangle$, and meanwhile we define $s=g_1^2+g_2^2$, $D=g_1^4+g_1^2g_2^2+g_2^4$ and $\delta u=u_1-u_2$. 
The first Lanczos step gives
\begin{equation}
\widetilde H_{\rm mm}|K_0\rangle
=
g_1|1,0\rangle+g_2|0,1\rangle,
\end{equation}
and hence we have $b_1=\sqrt{s}$ and $|K_1\rangle=
\frac{g_1|1,0\rangle+g_2|0,1\rangle}{\sqrt{s}}$. 
The first diagonal Lanczos coefficient is $a_1=\langle K_1|\widetilde H_{\rm mm}|K_1\rangle=\frac{2(u_1g_1^2+u_2g_2^2)}{s}$. Following the Lanczos algorithm, we could derive the second unnormalized residual which is defined by $|A_2\rangle
=\widetilde H_{\rm mm}|K_1\rangle-a_1|K_1\rangle-b_1|K_0\rangle$, its resulting formula is
\begin{align}
|A_2\rangle=\frac{2}{\sqrt{s}}\left(g_1^2|2,0\rangle
+g_1g_2|1,1\rangle+g_2^2|0,2\rangle\right)+\frac{2g_1g_2\delta u}{s\sqrt{s}}\left(g_2|1,0\rangle-g_1|0,1\rangle\right).
\label{A2-general}
\end{align}
The Eq.~\eqref{A2-general} belongs to the total-pair-number sectors $N_{\rm pair}=2$ and $N_{\rm pair}=1$.
They are mutually orthogonal. The corresponding Lanczos
coefficient is
\begin{equation}
b_2^2=\langle A_2|A_2\rangle=\frac{4D}{s}+\frac{4g_1^2g_2^2(\delta u)^2}{s^2}.
\label{b2-general}
\end{equation}
Thus, whenever $g_1g_2\delta u\neq0$, the normalized vector $|K_2\rangle=|A_2\rangle/b_2$ is not an eigenstate of the total pair-number operator. The  condition \eqref{number-Krylov-condition} is therefore already
violated at the second Lanczos level. This violation can be seen directly in the short-time expansion (Taylor expansion of time $t$). For $a_0=0$, the scalar Lanczos chain gives
\begin{equation}
C_K(t)
=
b_1^2t^2
+
\frac{b_1^2}{12}
\left(
2b_2^2-4b_1^2-a_1^2
\right)t^4
+
\mathcal O(t^6),
\label{CK-short-general}
\end{equation}
where we have utilized the $b_1=g$, $b_2=2g$ and $a_1=u_2$ in the single channel. On the other hand, Eq.~\eqref{multimode-N-definition} yields
\begin{align}
\left\langle \hat N_{\rm pair}\right\rangle_t=
(g_1^2+g_2^2)t^2+\frac{1}{3}\left[g_1^2(g_1^2-u_1^2)+g_2^2(g_2^2-u_2^2)\right]t^4+\mathcal O(t^6).
\label{N-short-general}
\end{align}
Substituting $a_1$ and
\eqref{b2-general} into Eq.~\eqref{CK-short-general}, we obtain
\begin{equation}
\boxed{
C_K(t)
-
\left\langle \hat N_{\rm pair}\right\rangle_t
=
\frac{
g_1^2g_2^2(u_1-u_2)^2
}{
g_1^2+g_2^2
}
t^4
+
\mathcal O(t^6)
}.
\label{detuning-mismatch}
\end{equation}
Equation~\eqref{detuning-mismatch} demonstrates that even without nonlinear interactions or non-Gaussian effects, Gaussian mode inequivalence is sufficient to generate Krylov information inaccessible to particle counting. The deviation is non-negative and vanishes in the single-channel and equal-detuning limits, meaning $u_1=u_2$.

\subsection{Breakdown induced by inequivalent pair-production channels}
\label{subsec:unequal-coupling}
We next isolate the effect of unequal pair-production amplitudes by considering the resonant limit, namely, 
\begin{equation}
u_1=u_2=0.
\end{equation}
In this case all diagonal Lanczos coefficients vanish because the Hamiltonian changes the total pair number by one unit and therefore alternates the pair-number parity. Equations of $b_1$ and $b_2$ reduce to $b_1=\sqrt{s}$ and $b_2=2\sqrt{\frac{D}{s}}$, and the second normalized Lanczos vector is
\begin{equation}
|K_2\rangle
=
\frac{
g_1^2|2,0\rangle
+
g_1g_2|1,1\rangle
+
g_2^2|0,2\rangle
}{\sqrt{D}}.
\label{K2-resonant}
\end{equation}
Although $|K_1\rangle$ and $|K_2\rangle$ have definite total pair
numbers one and two, respectively, the correspondence fails at the
next Lanczos step. The third unnormalized residual is
\begin{align}
|A_3\rangle
={}&
\frac{
g_1g_2(g_1^2-g_2^2)
}{
s\sqrt{D}
}
\left(
g_2|1,0\rangle-g_1|0,1\rangle
\right)
\nonumber\\
&+
\frac{3}{\sqrt{D}}
\Big[
g_1^3|3,0\rangle
+
g_1^2g_2|2,1\rangle
+
g_1g_2^2|1,2\rangle
+
g_2^3|0,3\rangle
\Big].
\label{A3-resonant}
\end{align}
The first line of Eq.~\eqref{A3-resonant} belongs to the
$N_{\rm pair}=1$ sector, whereas the second line belongs to the
$N_{\rm pair}=3$ sector. Therefore, we could derive $\hat N_{\rm pair}|K_3\rangle\not\propto |K_3\rangle$ with $g_1g_2(g_1^2-g_2^2)\neq0$. The corresponding coefficient is
\begin{equation}
b_3^2=\frac{9s(g_1^4+g_2^4)}{D}+\frac{g_1^2g_2^2(g_1^2g_2^2)^2}{sD}.
\label{b3-resonant}
\end{equation}
For a resonant tridiagonal Lanczos chain with $a_n=0$, the
short-time spread complexity through sixth order and using Eqs.~$b_1$, \eqref{K2-resonant}, and \eqref{b3-resonant}, Krylov complexity becomes as 
\begin{align}
C_K(t)
={}&
(g_1^2+g_2^2)t^2+
\frac{g_1^4+g_2^4}{3}t^4
\nonumber\\
&+
\left[
\frac{2(g_1^6+g_2^6)}{45}
+
\frac{
2g_1^2g_2^2(g_1^2-g_2^2)^2
}{
9(g_1^2+g_2^2)
}
\right]t^6
+
\mathcal O(t^8).
\label{CK-sixth-explicit}
\end{align}
Because the two channel Hamiltonians commute, the evolved state is a product of two two-mode squeezed states,
\begin{equation}
|\Psi(t)\rangle
=
\bigotimes_{\mu=1}^{2}
\left[
\frac{1}{\cosh(g_{\mu}t)}
\sum_{n_{\mu}=0}^{\infty}
\left(i\tanh(g_{\mu}t)\right)^{n_{\mu}}
|n_{\mu},n_{\mu}\rangle
\right],
\label{two-channel-state}
\end{equation}
up to an irrelevant convention-dependent phase. Hence, 
\begin{equation}
\left\langle \hat N_{\rm pair}\right\rangle_t
=
\sinh^2(g_1t)+\sinh^2(g_2t),
\label{N-resonant-exact}
\end{equation}
whose short-time expansion is $\left\langle \hat N_{\rm pair}\right\rangle_t=(g_1^2+g_2^2)t^2+\frac{g_1^4+g_2^4}{3}t^4+\frac{2(g_1^6+g_2^6)}{45}t^6+\mathcal O(t^8)$. 
Combining with Eq.~\eqref{CK-sixth-explicit}, we arrive at
\begin{equation}
\boxed{
C_K(t)
-
\left\langle \hat N_{\rm pair}\right\rangle_t
=
\frac{
2g_1^2g_2^2(g_1^2-g_2^2)^2
}{
9(g_1^2+g_2^2)
}
t^6
+
\mathcal O(t^8)
}.
\label{coupling-mismatch}
\end{equation}
The discrepancy first appears at order $t^6$ because the first two nontrivial Lanczos vectors still have definite pair numbers. The third vector is the first one that contains components from
different total-pair-number sectors, and its transition amplitude starts at order $t^3$, producing an order-$t^6$ contribution to the probability.

\subsection{Collective symmetric limit and physical interpretation}
\label{subsec:collective-limit}
The equality between Krylov complexity and pair number is recovered in two special limits. The first is the single-channel limit, $g_1g_2=0$. The second is the fully symmetric limit $g_1=g_2=g$ and $u_1=u_2=u$. In the latter case, the collective generators $K_a^{\rm col}=K_a^{(1)}+K_a^{(2)}$ with $a\in\{0,+,-\}$ form 
form a closed $SU(1,1)$ algebra, and
\begin{equation}
H_{\rm mm}=2uK_0^{\rm col}+
g\left(K_+^{\rm col}+K_-^{\rm col}\right).
\label{matrix element of hamiltonian}
\end{equation}
Starting from the two-channel vacuum, the scalar Lanczos evolution then remains inside a single collective $SU(1,1)$ representation. Its $n$-th collective ladder state is a superposition of mode distributions with a fixed total pair number $n$, so that we have $\hat N_{\rm pair}|K_n\rangle=n|K_n\rangle$ and the exact single-ladder identity is restored. 

Away from these fine-tuned limits, the scalar Krylov level counts the number of orthogonal dynamical directions generated by repeated actions of the full multimode Hamiltonian, whereas
$\hat N_{\rm pair}$ counts only the mean number of produced
correlated pairs. Equations~\eqref{detuning-mismatch} and
\eqref{coupling-mismatch} show that the difference between the two quantities is controlled by the inequivalence of the wave-packet channels. The Krylov construction is therefore sensitive not only to the total production yield but also to how the dynamics is distributed among distinct Gaussian pair-production channels. This provides the additional physical information that is absent in the closed  $SU(1,1)$ reduction.

\begin{figure}[htbp]
    \centering

    \begin{subfigure}{0.50\textwidth}
        \centering
        \includegraphics[width=\textwidth]{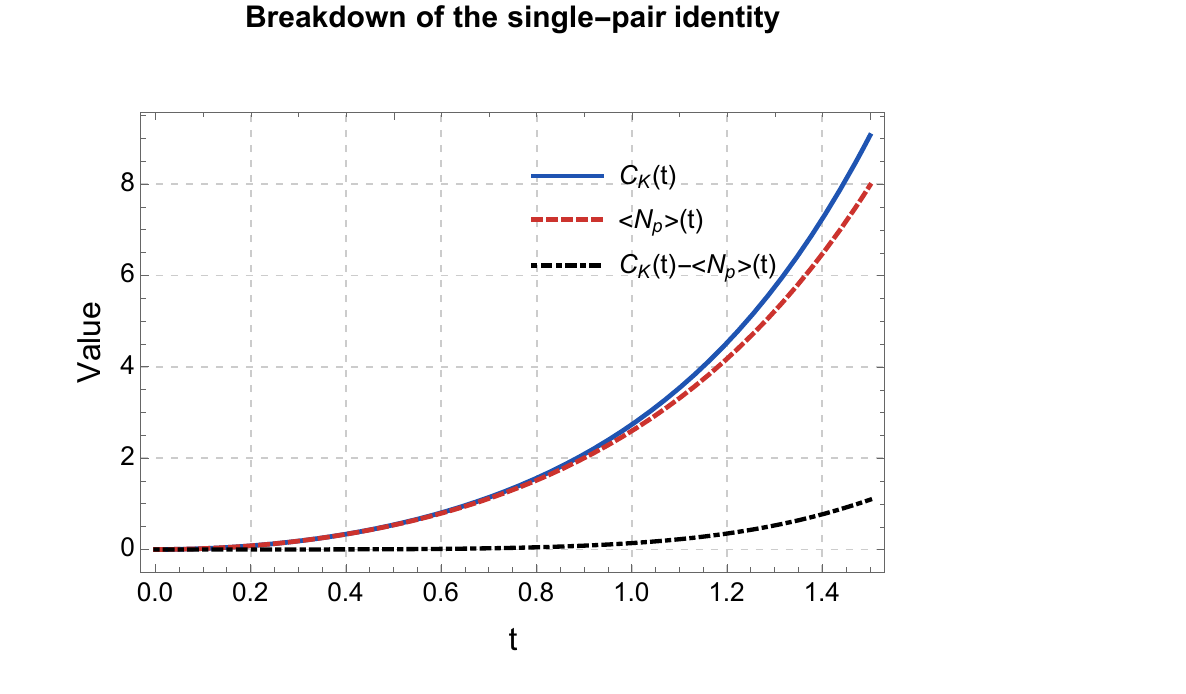}
        \label{fig:CK and pair}
    \end{subfigure}
    \hfill
    \begin{subfigure}{0.49\textwidth}
        \centering
        \includegraphics[width=\textwidth]{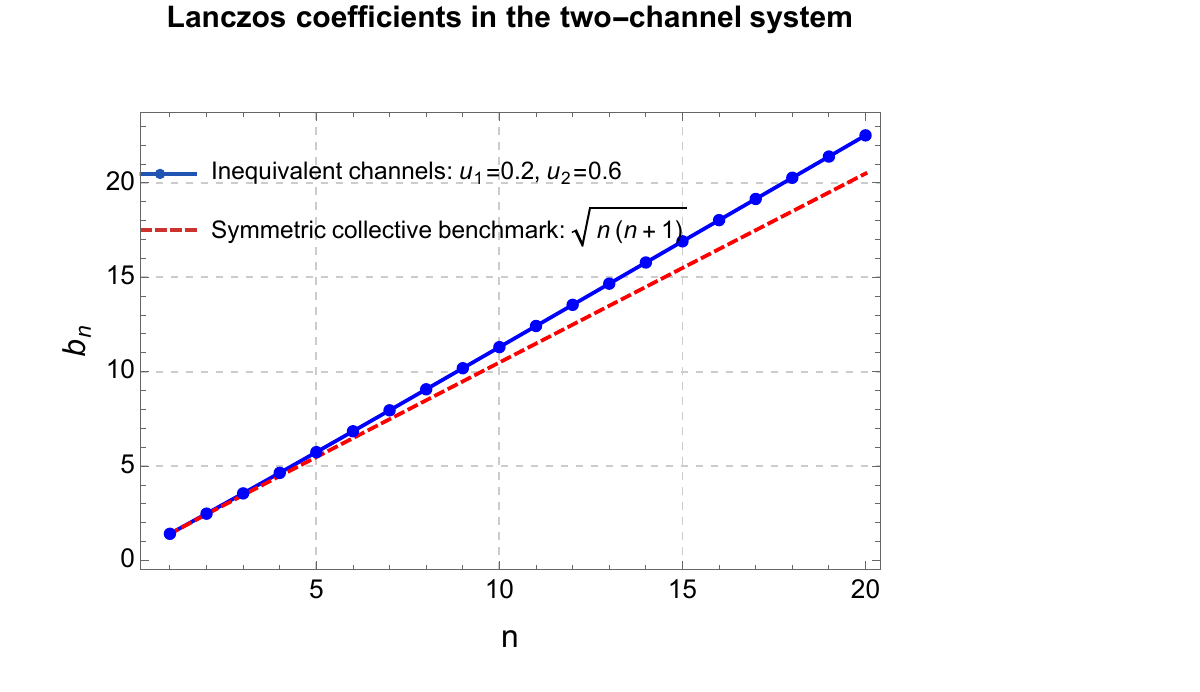}
        \label{fig:bn}
    \end{subfigure}

    \caption{\justifying The breakdown of  identiy and the Lanczos coefficient in the two-channel system. Left panel: Krylov complexity and mean-pair-number eigenstates within the short time interval. Right panel: The inequivalent vs symetric collective benchmark. The left panel adopt the $g_1=g_2=1$, $u_1=0.2$, and $u_2=0.6$, indicating that the Krylov complexity and the mean pair-number eigenstates will separate at the later times. The right panel shows $b_n$ for the same inequivalent two-channel system, compared with the symmetric collective-$SU(1,1)$ benchmark $b_n=\sqrt{n(n+1)}$ with $g_1=g_2=1$, $u_1=0.2$, and $u_2=0.6$, indicating the deviations for the large $n$.} 
    \label{fig:Ck and bn}
\end{figure}

To further demonstrate the physical distinction between Krylov spreading and particle production, we numerically implement the two-channel quadratic Hamiltonian in a truncated pair-number Hilbert space. Starting from the vacuum state $|K_0\rangle=|0,0\rangle$, we construct the Lanczos basis and evaluate both the Krylov complexity
$C_K(t)=\sum_n n|\phi_n(t)|^2$ and and the mean correlated pair number $\langle \hat N_{\rm pair}\rangle_t$. In the symmetric limit, the two quantities coincide, reflecting the
emergent collective $SU(1,1)$ structure. Away from this limit, the Lanczos vectors explore additional mode-distribution directions and $C_K(t)$ becomes larger than the mean pair-number eigenstates. Another key physical quantity is the Lanczos coefficient $b_n$. To evaluate this, we utilize the lowest weight representation of the $SU(1,1)$ algebra. We begin by defining the lowest weight state $\vert{}k,0\rangle$, which satisfies the relations $K_{-}\vert{}k,0\rangle=0$ and $K_{0}\vert{}k,0\rangle=k\vert{}k,0\rangle$, where $k$ represents the Bargmann index \cite{raczka1986theory}. Consequently, we obtain the compact formulas for the $SU(1,1)$ representation theory as follows: 
\begin{align}
K_0|k,n\rangle &= (k+n)|k,n\rangle ,
\\[2mm]
K_+|k,n\rangle &= 
\sqrt{(n+1)(2k+n)}\,|k,n+1\rangle ,
\\[2mm]
K_-|k,n\rangle &= 
\sqrt{n(2k+n-1)}\,|k,n-1\rangle .
\end{align}
where we have applied the symmetric conditions $g_1=g_2=g$ and $u_1=u_2=u$. This allows us to define collective generators, simplifying the system into a single $SU(1,1)$ algebra. In this context, the state $\vert{}k,n\rangle$ encompasses all modes under the symmetric limit; however, the generic case cannot be described by this single state. Based on the matrix elements of the Hamiltonian in Eq.~\eqref{matrix element of hamiltonian}, the Lanczos coefficient is explicitly given by: 
\begin{equation}
    b_n=|g|\sqrt{n(2k+n-1)},
    \label{lanczos coefficient}
\end{equation}
In our case, the symmetric collective benchmark corresponds to $\vert{}g\vert{}=1$ and $k=1$, yielding the Lanczos coefficient $b_n=\sqrt{n(n+1)}$. Under these conditions, we recover the exact relation $C_k=\vert{}\beta_k\vert{}^2$. To better illustrate the deviations from this standard formula, we present numerical results for the Krylov complexity $C_k$, the mean number density $\vert{}\beta_k\vert{}^2$, and the Lanczos coefficients $b_n$, as shown in Fig.~\ref{fig:Ck and bn}.

The left panel of Fig.~\ref{fig:Ck and bn} illustrates the deviation between $C_k$ and $\vert{}\beta_k\vert{}^2$. The blue line represents the complete evolution of the Krylov complexity based on its original definition, Eq.~\eqref{multimode-CK-definition}, utilizing the full Lanczos algorithm, while the red dashed line corresponds to Eq.~\eqref{multimode-N-definition}. These numerical results demonstrate the divergence between the Krylov complexity and the mean number density much more robustly than the short-time expansion in Eq.~\eqref{detuning-mismatch}, although the latter also analytically predicts this breakdown. In the right panel, we compare the Lanczos coefficients between the generic case and the symmetric limit. The numerical values of $b_n$ for the generic case are computed using the general recurrence relation of the Lanczos algorithm, 
\begin{equation}
 |A_{n+1}\rangle= H|K_n\rangle-a_n|K_n\rangle-b_n|K_{n-1}\rangle,
 \label{recurrence relation}
\end{equation}
with parameters $g_1=g_2=1$, $u_1=0.2$, and $u_2=0.6$ applied to the two-channel Hamiltonian \eqref{two-channel-H}. Specifically, the first few computed values are $b_1= 1.4142$, $b_2=2.4819$, and $b_3=3.5552$. Comparing these results with the symmetric limit reveals that deviations emerge for $n>2$. 

In this section, we have clearly demonstrated that the relation $C_k=\vert{}\beta_k\vert{}^2$ is no longer valid for the multi-mode case, even within the quadratic Hamiltonian \eqref{two-channel-H}. This relation holds only in the symmetric limit where $u_1=u_2$ and $g_1=g_2$. Analytically, using a short-time expansion, Eqs. \eqref{detuning-mismatch} and \eqref{coupling-mismatch} clearly reveal this breakdown across various detuning parameters and distinct coupling constants. To fully capture these deviations, we provide a numerical comparison between $C_k$ and $\langle N_k \rangle=\vert{}\beta_k\vert{}^2$ using the general Lanczos algorithm, as shown in the left panel of Fig. \ref{fig:Ck and bn}. Furthermore, we utilized the general recurrence relation \eqref{recurrence relation} to numerically evaluate $b_n$, comparing it to the symmetric limit (where $C_k=\vert{}\beta_k\vert{}^2$ remains valid). Our numerical results show that a deviation emerges for $n>2$. Finally, we could see that only the single (including the symmetric limit) $SU(1,1)$ could hold the formula $C_K=|\beta_k|^2$.

\section{Conclusion and Outlook}
\label{conclusion and outlook}

In this work, we have established a comprehensive observer-dependent Krylov framework tailored for non-inertial quantum systems. A central conceptual pillar of this framework is that the Minkowski vacuum can be elegantly expanded in terms of Bogoliubov coefficients, as shown in Eq.~\eqref{genera wave function}, naturally rendering the Rindler pair-number sectors as the Krylov basis. By generalizing this Bogoliubov evolution under the effective $SU(1,1)$ Hamiltonian \eqref{total hamiltonian}, we arrive at a profoundly simple and exact identity: the Krylov complexity is strictly equivalent to the mean number of Rindler pairs. Consequently, operator growth within Krylov space acquires a tangible, relativistic interpretation—namely, it strictly quantifies the dynamic production and spreading of correlated Rindler pairs.

The dynamics are controlled by the competition between pair production and detuning. In terms of the dimensionless parameter $\eta=u_2/g$, the system exhibits three distinct regimes, as shown in Fig.~\ref{fig:comparison}. For $|\eta|<1$, the Krylov wave packet spreads hyperbolically along the pair-number chain. At $|\eta|=1$, the growth becomes critical. For $|\eta|>1$, the evolution becomes bounded, and the probability distribution remains exponentially concentrated near low Krylov levels. This provides a clear probability-level mechanism for detuning-induced Krylov localization, as presented in Fig.~\ref{fig:distribution}. However, the relation $C_k=\vert{}\beta_k\vert{}^2$ holds only for a single $SU(1,1)$ Hamiltonian or in the symmetric limit, as shown in Eq.~\eqref{matrix element of hamiltonian}. It breaks down in the multi-mode case, such as the two-channel Hamiltonian \eqref{two-channel-H}. To illustrate this violation, we perform a short-time expansion of the Krylov complexity, incorporating different detuning parameters and various coupling constants. Our analytical derivations explicitly confirm this breakdown. Furthermore, using the general Lanczos algorithm, Fig.~\ref{fig:Ck and bn} clearly demonstrates the deviation between the standard symmetric case and the generic case. 

This framework admits a natural extension to more general scenarios, encompassing broader group-theoretic structures of the underlying Hamiltonian \cite{Caputa:2021sib}. A particularly interesting direction is to investigate interacting quantum fields in curved spacetime. While particle production and Bogoliubov mixing have been extensively studied in dynamical gravitational backgrounds \cite{Davies:1974th,Fulling:1972md}, a Krylov description of such systems remains largely unexplored. In this setting, Krylov spreading may encode dynamical information associated with gravitational backreaction, nonlinear mode coupling, and operator growth beyond Gaussian particle production. From a relativistic standpoint, these generalizations hold significant potential for applications in quantum information channels in curved spacetime, moving-cavity systems, and relativistic quantum metrology \cite{Bruschi:2012uf,Ahmadi:2013pfa,Barcellos:2023kpi}. Conversely, within the Krylov domain, it would be highly compelling to integrate our current pair-number chain formulation with open-system Krylov complexity, spread complexity, multi-seed Krylov formalisms, and entanglement-based diagnostics of operator growth \cite{Liu:2022god,Bhattacharya:2022gbz,Balasubramanian:2022tpr,Caputa:2022yju,Patramanis:2021lkx,Craps:2024suj}. Together, these future directions promise to elucidate how information scrambling, Krylov localization, and quantum processing manifest when the non-inertial motion of the observer is intrinsically involved into the quantum-field-theoretic fabric.

\acknowledgments
M.-Q. Ma, S.-C. Liu and L.-H. Liu are funded by NSFC grant NO. 12665009 and  Hunan Provincial Department of Education Project NO. 25B0480; H.-Q. Zhang was partially supported by NSFC with grant NO. 12675057.

\bibliography{Refs}
\end{document}